\documentclass[twocolumn, floatfix]{aastex631}

\usepackage{float}
\usepackage{gensymb}
\usepackage[normalem]{ulem}
\usepackage{multirow}
\usepackage{longtable}
\usepackage{threeparttable} 
\usepackage{hyperref}
\usepackage{amsmath} 

\usepackage{xcolor}
\usepackage{breqn}
\usepackage[flushmargin]{footmisc}
 \usepackage{booktabs}
\usepackage{tikz}
\usetikzlibrary{calc,tikzmark} 
\newlength{\imageheight}

\shorttitle{Typhon Stellar Stream}
\shortauthors{Nogueira-Santos et al.}

\graphicspath{{./}{figures/}}

\begin{document}

\title{Evidence for the Typhon Stellar Stream as the Remnant of a Recently-disrupted Dwarf Galaxy}

\correspondingauthor{J. V. Nogueira-Santos}
\email{joao.nogueira@inpe.br}

\author[0009-0007-5867-0583]{Jo\~ao~V.~Nogueira-Santos}
\affiliation{Instituto Nacional de Pesquisas Espaciais, Av. dos Astronautas 1758, Jardim da Granja, 12227-010 S\~ao Jos\'e dos Campos, SP, Brasil}
\affiliation{Universidade de S\~ao Paulo, Instituto de Astronomia, Geof\'isica e Ci\^encias Atmosf\'ericas, Departamento de Astronomia, SP 05508-090, S\~ao Paulo, Brasil}


\author[0000-0002-9269-8287]{Guilherme Limberg}
\affiliation{Kavli Institute for Cosmological Physics, University of Chicago, 5640 S. Ellis Avenue, Chicago, IL 60637, USA}
\affiliation{Department of Astronomy \& Astrophysics, University of Chicago, 5640 S. Ellis Avenue, Chicago, IL 60637, USA}

\author[0000-0001-7479-5756]{Silvia~Rossi}
\affiliation{Universidade de S\~ao Paulo, Instituto de Astronomia, Geof\'isica e Ci\^encias Atmosf\'ericas, Departamento de Astronomia, SP 05508-090, S\~ao Paulo, Brasil}

\author[0000-0002-0537-4146]{H\'elio~D.~Perottoni}
\affiliation{Observat\'orio Nacional, MCTI, Rua Gal. Jos\'e Cristino 77, Rio de Janeiro, 20921-400, RJ, Brasil}

\author[0000-0003-3382-1051]{Lais~Borbolato}
\affiliation{Universidade de S\~ao Paulo, Instituto de Astronomia, Geof\'isica e Ci\^encias Atmosf\'ericas, Departamento de Astronomia, SP 05508-090, S\~ao Paulo, Brasil}

\author[0000-0002-8262-2246]{Fabr\'icia~O.~Barbosa}
\affiliation{Universidade de S\~ao Paulo, Instituto de Astronomia, Geof\'isica e Ci\^encias Atmosf\'ericas, Departamento de Astronomia, SP 05508-090, S\~ao Paulo, Brasil}

\author[0000-0002-7758-656X]{Andr\'e~R.~da~Silva}
\affiliation{Universidade de S\~ao Paulo, Instituto de Astronomia, Geof\'isica e Ci\^encias Atmosf\'ericas, Departamento de Astronomia, SP 05508-090, S\~ao Paulo, Brasil}

\begin{abstract}

In the hierarchical model of galaxy formation, the Milky Way’s stellar halo assembles through the successive accretion of dwarf galaxies. While many resulting substructures have been identified, the origin of the recently discovered Typhon stellar stream remains ambiguous, as dynamical data alone are insufficient to distinguish between a globular cluster and a dwarf galaxy progenitor. To resolve this, we combine astrometry from Gaia with spectroscopic data from 
available surveys to comprehensively analyze the stream's chemical and dynamical properties. We identify 107 candidate members, significantly expanding the known sample. Three of these candidates lie beyond $55\,\mathrm{kpc}$, with the farthest at $67\,\mathrm{kpc}$, lending further support to Typhon’s inferred large apocentric distance. Our analysis yields a mean metallicity of $\rm[Fe/H]=-1.42\pm0.07$ and a metallicity dispersion of $\sigma_{\text{[Fe/H]}} = 0.26^{+0.06}_{-0.05}$\,dex. This metallicity dispersion supports a dwarf-galaxy origin for the Typhon stream. Typhon’s high orbital energy is consistent with a relatively recent accretion of a fully disrupted dwarf galaxy with a stellar mass in the range of $10^6 M_{\odot} < M_\star \lesssim 10^7 M_{\odot}$. Furthermore, Typhon is the only major dwarf-galaxy stellar stream in the Milky Way's halo to exhibit such an extreme orbital eccentricity of $e \gtrsim 0.85$. We conjecture that dwarf-galaxy streams from high-eccentricity mergers are fully dissolved much more rapidly than those on circular orbits, in line with Typhon's proposed recent accretion. Additional dwarf-galaxy stellar streams on high-eccentricity orbits should be discoverable with next-generation photometric surveys as well as future releases from the Gaia space mission.


\end{abstract}

\keywords{Milky Way stellar halo; Stellar streams; Milky Way dynamics; Dwarf galaxies; Metallicity}

\defcitealias{Tolstoy2023}{T23}
\defcitealias{delosReyes22}{R22}

\section{Introduction} \label{sec:intro}
\setcounter{footnote}{0}

Within the hierarchical formation paradigm, the Galactic stellar halo is expected to be assembled primarily through the continuous accretion and merging of smaller satellite systems \citep{sz1978, white1978,White1991cdm,Johnston1998}. When a smaller galaxy falls into a massive host, the intense gravitational gradient 
exerts severe tidal forces on the infalling system \citep{HelmiWhite1999}. These forces pull stars away from the progenitor, stretching them along the orbital path to form a stellar stream \citep[e.g.,][for a review]{Bonaca2024streamsREVIEW}. As spatially coherent ribbons of stellar debris, these structures, representing an intermediate phase in the tidal dissolution of either dwarf galaxies or globular clusters, resist immediate phase-mixing and retain the kinematic signatures of their progenitor's orbit \citep{BullockJohnston2005, Cooper2010, Pillepich2015halos}.

Thanks to Gaia's astrometric data \citep{GaiaMission}, the number of stellar streams escalated to nearly a hundred \citep{Mateu2022galstreams}
. This census is, however, largely biased toward nearby systems within $\sim$20\,kpc from the Sun \citep[][]{Malhan2022atlas}, which is an effect of Gaia's 
magnitude limit. These inner portions of the Galactic halo are mostly 
associated with ancient accretion events that had enough time to sink deep into the Milky Way's
potential via dynamical friction \citep{Rocha2012}. Differently, the debris of recently accreted systems should be found in the outermost regions of the Galactic halo \citep{Horta2023haloSubs}. Although a few streams are known beyond 30\,kpc \citep{Grillmair2009, Shipp2018}, stars (including red giant-branch and horizontal-branch stars) are so faint at such distances that acquiring spectra 
is quite time consuming even for 6--10\,m-class telescopes.

In June 2022, a peculiar stellar stream was discovered in Gaia's third Data Release \citep[DR3;][]{Gaiadr3}
data, dubbed ``Typhon'' \citep{Tenachi2022typhon}. The orbits of its stars have apocenters reaching ${\gtrsim}$100\,kpc. Nevertheless, Typhon stars have been found crossing the Solar neighborhood, close to their pericenters (${\sim}4$\,kpc). Therefore, the Typhon stream provides the unique opportunity for us to have a glimpse at the properties of the outer Galactic halo with nearby, hence reasonably bright, stars. 

Right after its discovery, the abundance patterns of seven Typhon stars were studied with high-resolution spectroscopy \citep{AlexJi2023typhon}. The system appears to have a significant metallicity dispersion, indicative of a dwarf-galaxy origin \citep{Willman2012galaxyDefined}, but this result relies on the membership of a single metal-poor star at $\rm[Fe/H] < -2$. 
The rest of the known Typhon members are within $-1.7 < \rm[Fe/H] \lesssim -1.5$,  which would be consistent with no metallicity spread after accounting for uncertainties, hence a globular cluster progenitor. Given the small size of the analyzed sample, additional chemical information is certainly needed to confirm the origin of this substructure.

This work is organized as follows. In Section \ref{sec:data}, we describe the data. In Section \ref{sec:methods}, we outline the methods employed for orbit integration and candidate selection. We present our results and discussions in Section \ref{sec:res} and conclude with a summary in Section \ref{sec:conc}.

\section{Data} \label{sec:data}
In this work, we utilize data from several spectroscopic surveys, which were combined with astrometric data from Gaia~DR3 and spectro-photometric distances from the literature. For consistency across these heterogeneous datasets, we apply a basic set of quality cuts. Unless otherwise noted, we restrict all samples to stars with effective temperatures in the range $4000 < T_{\mathrm{eff}}/{\rm K} < 7000$. Furthermore, we remove sources with potentially spurious astrometric solutions by enforcing an astrometric fidelity threshold of $\mathtt{fidelity\_v2} \geq 0.5$ \citep{Rybizki2022fidelity} and  exclude known variable stars by requiring $\mathtt{varFlag} == \mathrm{'NOT\_AVAILABLE'}$ \citep{GaiaDR32022arXiv}. Crucially, to isolate the stellar stream from the foreground metal-rich Milky Way population, we limit our analysis to metal-poor stars with $[\mathrm{Fe/H}] \leq -0.6$. This threshold minimizes contamination from the Galactic disk without compromising the recovery of Typhon members. Because this work combines heterogeneous spectroscopic datasets, we chose not to apply a single, uniform signal-to-noise ratio ($S/N$) threshold. Instead, we adopt survey-specific  cuts tailored to the spectral resolution and pipeline characteristics of each project.  

\subsection{SEGUE}

The Sloan Extension for Galactic Understanding and Exploration (SEGUE), conducted using the 2.5-m telescope at the Apache Point Observatory/USA, is a spectroscopic survey that targeted distant metal-poor stars in the Galactic halo \citep[][]{Yanny2009sgr, Rockosi2022segue} and has been extensively utilized in Galactic Archaeology \citep[e.g.,][]{IvezicBeersJuric2012}. The SEGUE DR12 dataset used in this work provides stellar atmospheric parameters derived by the SEGUE Stellar Parameter Pipeline \citep[][]{seguesspp1, seguesspp2} from low-resolution ($R \sim 1800$) spectra of over 400,000 stars. We require \(S/N>10\) for the selected sample.

\subsection{LAMOST}

The Large Sky Area Multi-Object Fiber Spectroscopic Telescope (LAMOST), located at the Xinglong Station of the National Astronomical Observatory of China 
operates with a limiting magnitude of $r \sim 18$ and a spectral resolution of $R \sim 1800$ \citep{Cui2012, Liu2014}, similar to SEGUE. The LAMOST DR11 has collected over 10 million spectra spanning the wavelength range of $3690–9100\,\mathrm{\AA}$. This dataset includes a catalog of over 9 million stars with atmospheric parameters derived via the LAMOST Stellar Parameter Pipeline \citep{LamostPipeline2015, LamostPipeline2018, LamostPipeline2021}. We require a minimum \(S/N\) of 15 across the \(g\), \(r\), \(i\), and \(z\) bands for the selected sample.

\subsection{DESI}

The Dark Energy Spectroscopic Instrument (DESI) is a moderate-resolution spectrograph ($R \sim 5000$ over $3600 < \lambda/\mathrm{\AA} < 9800$) mounted at the prime focus of the Mayall 4-m telescope located at Kitt Peak National Observatory/USA \citep{desi}. Although its primary objective is to constrain cosmological parameters, DESI also executes a Milky Way Survey component 
targeting approximately 7 million stars down to a limiting magnitude of $r \sim 19$ \citep{desimws}. This program specifically probes the Galactic stellar halo, yielding a dense spatial sampling of faint stars that is optimal for identifying substructures \citep[see][]{Kizhuprakkat2026subsDESI}. For this analysis, we utilize the DESI DR1 stellar catalog of radial velocities, abundances, and atmospheric parameters, which were derived using the $\tt{RVSpecFit}$ pipeline \citep{koposov2025}. We require a minimum \(S/N\) of 10 across the $B$, $R$, and $Z$ bands for the selected sample.

\subsection{Gaia XP and RVS }

In this work, we initially use radial velocities from the Gaia Radial Velocity Spectrometer \citep[RVS;][]{GaiaRVS} to identify kinematic members of the Typhon stream. The RVS is a near-infrared ($845–872$\,nm), medium-resolution ($R \sim 11{,}500$) slitless spectrograph that provides radial velocities for stars down to a limiting magnitude of $G_{\mathrm{RVS}} \sim 14$. Subsequently, we obtained metallicities for these candidates from the data-driven Gaia~XP-based catalog of stellar parameters from \citet{Andrae2023}. The Gaia~XP catalog provides low-resolution spectra ($R \sim 20-100$) for over 220 million sources, spanning the optical to near-infrared range \citep[$3300–10500$\,\AA;][]{DeAngeli2023}. Despite this low resolution, data-driven methods have successfully derived stellar parameters with high fidelity. However, we note that metallicity precision degrades at fainter magnitudes due to increased noise, particularly in the blue spectral region. We require 
a minimum $S/N\geq2$ in RVS spectra 
for the selected sample. With our criteria, five previously identified members of the Typhon stream were excluded \citep{Tenachi2022typhon}; three due to missing data in the Gaia~XP catalog, and the remaining two because they were classified as variable stars.

\subsection{Distances} 

With the exception of the DESI sample, for which we utilize the survey's native spectro-photometric distance estimates \citep{LiSongting2025desiDists}, we adopt the distances from \citet{Queiroz2023starhorse} derived via the {\tt StarHorse} code \citep{Santiago2016starhorse, Queiroz2018}. {\tt StarHorse} is a Bayesian isochrone-matching method that obtains distances, extinctions, and other parameters based on a set of observables and priors by comparing available data to stellar evolutionary models from the PAdova and TRieste Stellar Evolution Code \citep[PARSEC;][]{bressan2012parsec}. We use this algorithm, which has been extensively validated and applied to the outer Galactic halo in previous studies \citep{Perottoni2022gse, Limberg2023sgr}, because Gaia parallaxes alone are often insufficient for accurately determining distances to distant stars 
$({\gtrsim}10 \ {\rm kpc})$ within the Galactic halo. Furthermore, we apply distance quality cuts for each survey sample. We adopt a relative distance uncertainty cut where the distance error represents less than 50\% of the measured distance, to maximize the number of selected stars, and a stricter cut of less than 20\% to define a more reliable sample.

\section{Methods} \label{sec:methods}

\subsection{Kinematics, dynamics and member selection }

\begin{figure}[t!]
\begin{center}
\includegraphics[width=\columnwidth]{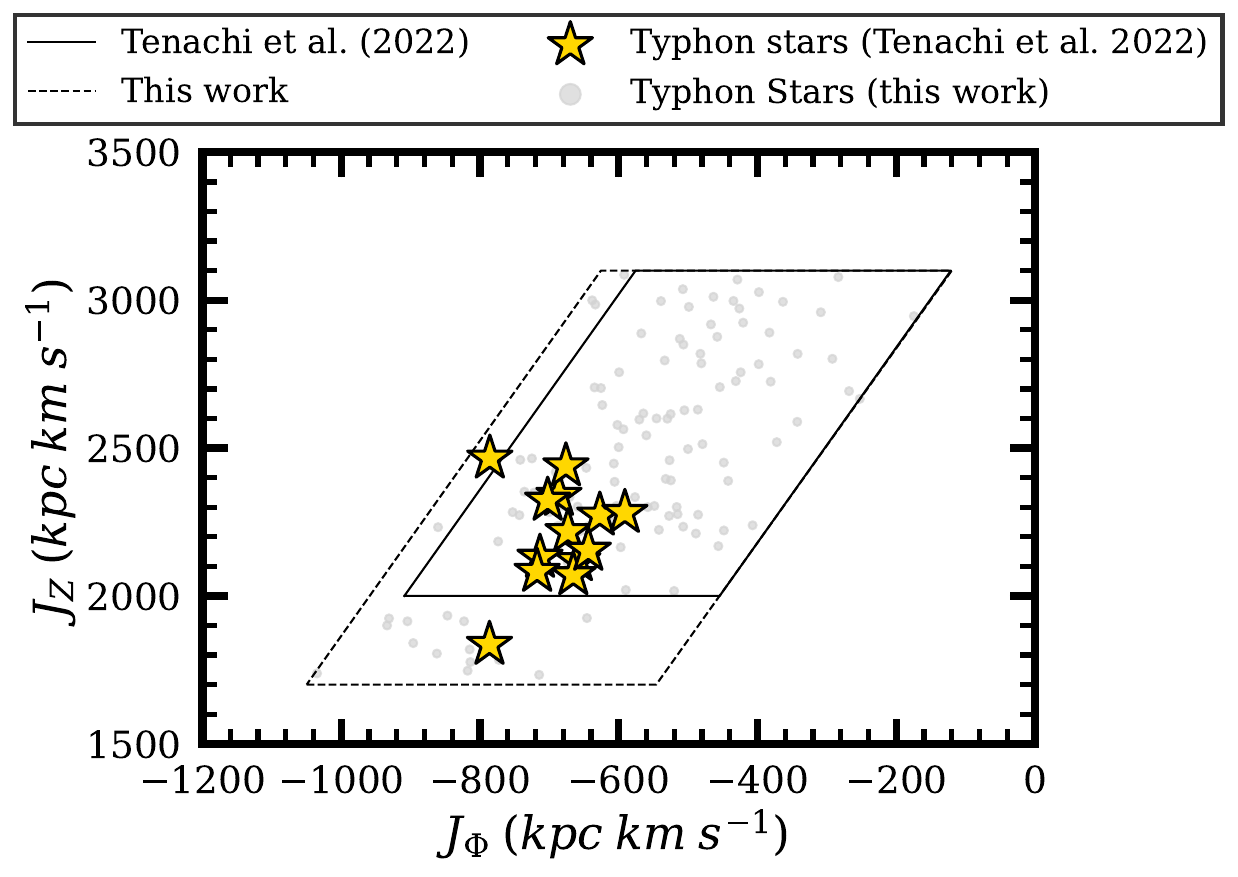}
\caption{Definition of the extended Typhon selection box in the $J_\Phi$ vs. $J_z$ space. Previously known Typhon members from \cite{Tenachi2022typhon} that were not excluded by the quality cuts outlined in Section \ref{sec:data} are shown as gold stars, with their orbital actions computed using our catalog parameters. The solid black parallelogram denotes the original dynamical boundaries defined by \cite{Tenachi2022typhon}. The dashed lines highlight the extended selection region adopted in this work to accurately enclose all the candidates. The background grey points represent the additional stars in our final sample that were not identified in the original \cite{Tenachi2022typhon} study}
\label{fig:box_selection}
\end{center}
\end{figure}

Using the {\tt AGAMA} package \citep{agama}, we integrate orbits forward in time for 20\,Gyr within the \cite{mcmillan2017} axisymmetric model potential. To establish the reference frame for these integrations, we set the solar coordinates to $(x,y,z)_{\odot} = (-8.2240, 0.0000, 0.0028)\,{\rm kpc}$, utilizing the solar radius derived by \cite{Bovy2020} and the vertical displacement from \cite{Widmark2021Sun}. Additionally, the solar velocity is defined as $(v_x, v_y, v_z)_{\odot} = (11.10, 250.20, 7.25)\,{\rm km \,s^{-1}}$, which includes both the peculiar motion of the Sun \citep{schon2010} and the local circular velocity of $243\,{\rm \,km\,s^{-1}}$ \citep{Bovy2020}.

To account for observational errors, we employ a Monte Carlo scheme with 100 realizations, in which the final orbital parameters are defined as the medians of the resulting distributions, adopting the 16th and 84th percentiles as uncertainties. In action space, a star's orbit is characterized by the action vector in cylindrical coordinates $\textbf{J} = (J_R, J_\phi, J_z)$. In this framework, $J_R \in  [0, \infty]$ quantifies the radial oscillations of the orbit and is related to orbital eccentricity, $J_\phi \in [-\infty, \infty]$ corresponds to the angular momentum around the Galactic axis of symmetry and is equivalent to the vertical component of angular momentum ($L_z$) in a Cartesian frame, and $J_z \in [0, \infty]$ measures the oscillations around the equatorial plane, i.e., its vertical excursion with respect to the Galactic plane \citep{binney2012}. Because these actions are fully conserved within an axisymmetric potential, stars belonging to the same stellar stream will naturally cluster together in this space.

\begin{table*}[t]
    \centering
    \caption{Number of identified Typhon stellar stream candidates, mean metallicity, and metallicity dispersion for each astronomical survey for distance errors representing less than 20\% and 50\% of the measured distance.}
    \label{tab:typhon_counts}
    \begin{tabular}{lcccccc}
    \toprule
    & \multicolumn{3}{c}{$\sigma_d/d < 50\%$} & \multicolumn{3}{c}{$\sigma_d/d < 20\%$} \\
    \cmidrule(lr){2-4} \cmidrule(lr){5-7}
    Survey & $N$ & $\langle \mathrm{[Fe/H]} \rangle$ & $\sigma_{\mathrm{[Fe/H]}}$ & $N$ & $\langle \mathrm{[Fe/H]} \rangle$ & $\sigma_{\mathrm{[Fe/H]}}$ \\
    \midrule
    DESI        & 72 & $-1.44 \pm 0.02$ & $0.10 \pm 0.09$ & 24 & $-1.44 \pm 0.03$ & $0.09 \pm 0.15$ \\
    SEGUE       & 9  & $-1.61 \pm 0.15$ & $0.45 \pm 0.13$ & 9  & $-1.61 \pm 0.15$ & $0.45 \pm 0.13$ \\
    LAMOST      & 19 & $-1.42 \pm 0.07$ & $0.26 \pm 0.09$ & 16 & $-1.41 \pm 0.08$ & $0.28 \pm 0.10$ \\
    Gaia RVS/XP & 19 & $-1.41 \pm 0.04$ & $0.15 \pm 0.11$ & 19 & $-1.41 \pm 0.04$ & $0.15 \pm 0.11$ \\
    \bottomrule
    \end{tabular}
\end{table*}
We first identify Typhon candidates by selecting only stars with orbital apocenters $>$70\,kpc and using a polygon selection based on the dynamical properties of the previously known members described by \cite{Tenachi2022typhon}. Figure \ref{fig:box_selection} displays these previously identified Typhon members, with their orbital actions calculated using the parameters from our current catalog. The original selection boundaries defined by \citet[][solid black lines]{Tenachi2022typhon} fail to recover these known members when their action vectors are computed with our data. This discrepancy is likely attributed to the distance estimates adopted in the original study, which were not explicitly reported. To overcome this limitation and ensure a complete recovery of the known members, we establish an extended Typhon selection. This extended region (indicated by the dashed lines) is defined by the boundaries $3100 < J_z/({\rm kpc} \,{\rm km}\,{\rm s}^{-1}) < 2000$ and $-120 < J_\phi/({\rm kpc} \,{\rm km}\,{\rm s}^{-1}) < -1050$, isolating the specific locus in the $J_z -J_\phi$ plane where the Typhon stars are located.

Because stellar streams are kinematically coherent structures, identifying a sample of stars in the same orbital phase allows us to determine the full stream orbit through integration. We evaluate Typhon's orbit from its members in the solar neighborhood, where a higher availability of measured radial velocities yields a large number of confirmed members, enabling the detection of stars at various orbital positions along the structure. From the sample obtained via the aforementioned box selection, we isolate stars within this region that exhibit velocity vectors similar to those identified by \cite{Tenachi2022typhon}, as they share a similar orbital phase. Given that orbits are conserved among stars in a coherent stream, we compute the orbits for all selected stars within 4\,kpc from the Sun and then identify candidate members beyond this region by selecting those with velocity vectors tangential to the mean Typhon orbit.

\section{Results and Discussions} \label{sec:res}
\subsection{Member candidates}
To robustly determine the origin of the Typhon stellar stream, it is essential to analyze a large-enough sample to map its full metallicity distribution function. Allowing for up to 50\% fractional distance uncertainty, we identify 9 members from SEGUE, 19 from LAMOST, 72 from DESI, and 19 from Gaia RVS/XP, as shown in Table \ref{tab:typhon_counts}. It should be noted that low-metallicity stars from the DESI survey are mostly excluded from this count due to their high distance uncertainties (see Appendix). Cross-matching these subsamples reveals minor overlaps: Gaia RVS/XP shares 6 stars with LAMOST and 1 with DESI, while LAMOST shares 1 star with SEGUE and 4 with DESI. Throughout this work, the LAMOST sample is utilized to illustrate the selection process in the main figures. However, identical procedures are applied to the other surveys. The corresponding figures and selected stars for the remaining samples are provided in the Appendix.

\begin{figure}[t!]
    \centering
    \includegraphics[width=0.45\textwidth]{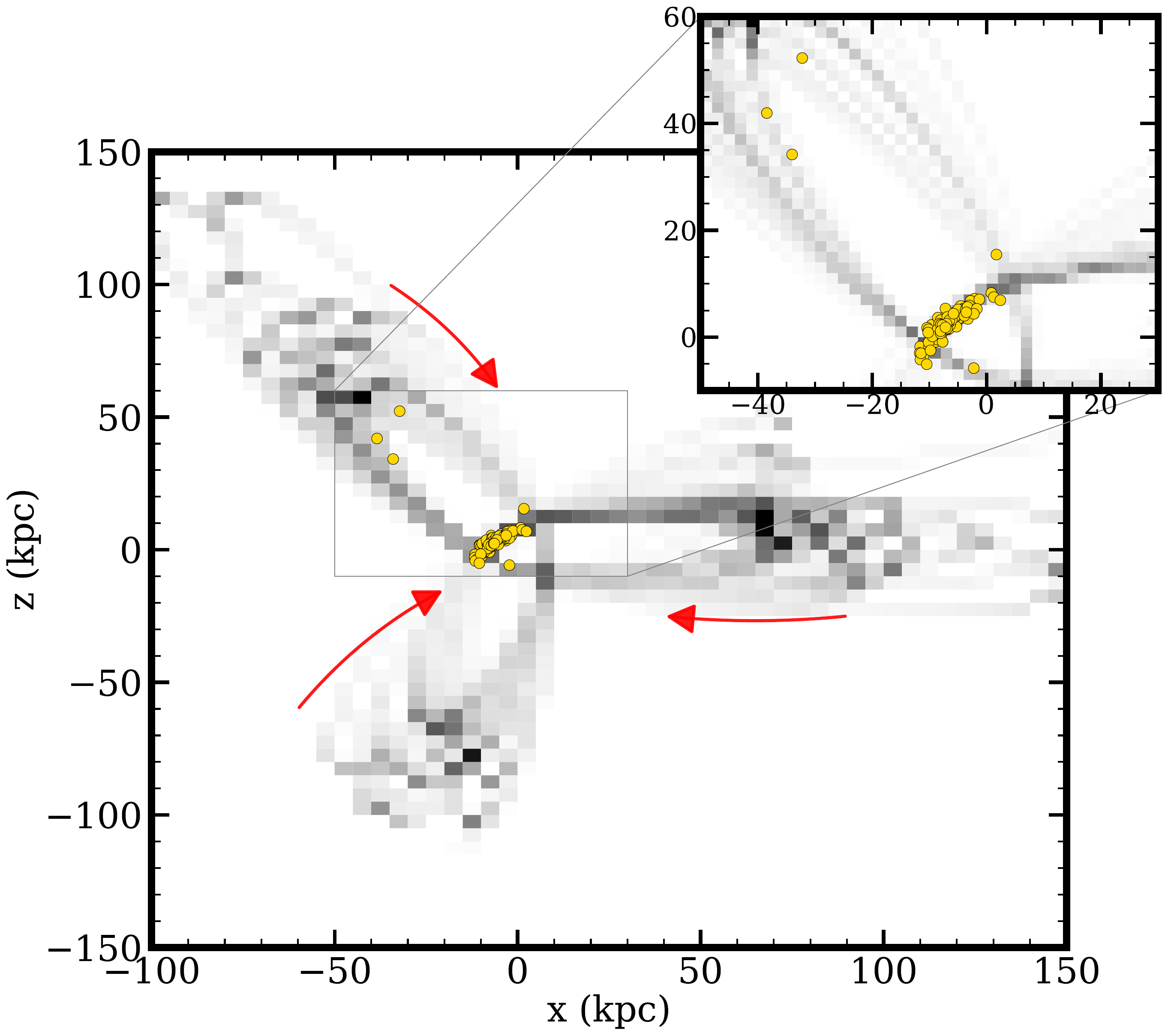}
    \hfill
    \caption{Spatial distribution of Typhon candidates. The main panel displays the candidate stars (yellow points) projected onto the Cartesian $x-z$ plane, overlaid on a grayscale density histogram representing the orbital tracks of the reference subgroup identified in the solar neighborhood. The inset panel provides a zoomed-in view of the spatial concentration of the identified members. Red arrows indicate the predicted velocity vectors along the stream's path at various orbital phases.}
    \label{fig:xz}
\end{figure}

Figure \ref{fig:xz} illustrates the spatial distribution of these 107 identified Typhon candidates, denoted by yellow points. Typhon's orbit passes close to the solar neighborhood at its pericenter, while its apocenter extends deep into the outer Milky Way halo, reaching $>$100\,kpc from the Galactic center. We observe a strong concentration of candidate stars near the Sun's position. This clustering reflects an observational selection effect (closer objects are systematically brighter and easier to detect) rather than an intrinsic physical absence of stream members in other orbital phases. Additionally, the red arrows indicate the predicted velocity direction of Typhon stars at various stages of the orbit. Overall, this spatial distribution explicitly reveals a cohesive filamentary structure characteristic of a stellar stream.

For a stellar stream with an extremely high apocenter, such as Typhon, stars are expected to exist across different orbital phases and distances. However, such distant members had not been previously identified. Although the majority of the selected Typhon stars are located in the solar neighborhood, defined in this work as the region at heliocentric distances $d < 4$ \,kpc, we identify three candidate members in the DESI survey at distances greater than 55\,kpc from the Sun, with the most distant one at 67\,kpc from the Sun
. This strongly suggests that Typhon's high apocenter is an intrinsic orbital feature rather than an observational artifact. Moreover, finding stars in different orbital configurations is crucial for reliably modeling Typhon’s kinematics and dynamics. This modeling could prove relevant for estimating the Milky Way's gravitational potential and provides important indications about the system's accretion history, since stars in different orbital phases were likely stripped at different times, as is the case for the Sagittarius dwarf-galaxy stream \citep{Ibata1994, Majewski2003}, with those in the satellite's outskirts becoming unbound earlier \citep[e.g,][]{Vasiliev2021tango, Limberg2023sgr, Cunningham2024sgr, almeidafernandes2026sgr}. 

\begin{table*}
\centering
\caption{Orbital and kinematic parameters for the analyzed objects. The columns represent the radial, azimuthal, and vertical actions ($J_r$, $J_\phi$, $J_z$), pericentric and apocentric distances ($r_{\rm peri}$, $r_{\rm apo}$), eccentricity ($e$), and total energy ($E_{\rm tot}$).}
\label{tab:parametros_orbitais}
\small 
\setlength{\tabcolsep}{5pt}
\renewcommand{\arraystretch}{1.0} 
\begin{tabular}{l c c c c c c c} 
\hline \hline
Object & $J_r$ & $J_\phi$ & $J_z$ & $r_{\rm peri}$ & $r_{\rm apo}$ & $e$ & $E_{\rm tot}$ \\
       & (kpc km s$^{-1}$) & (kpc km s$^{-1}$) & (kpc km s$^{-1}$) & (kpc) & (kpc) & & (km$^2$ s$^{-2}$) \\
\hline
Typhon & $5330 \pm 830$ & $-570 \pm 160$ & $2480 \pm 360$ & $6.4 \pm 1.1$ & $86 \pm 11$ & $0.86 \pm 0.04$  & $-69400 \pm 5300$ \\
GSE & $1390 \pm 390$ & $10 \pm 250$ & $480 \pm 750$ & $1.4 \pm 2.2$ & $20.8 \pm 7.3$ & $0.89 \pm 0.07$ & $-136000 \pm 16000$ \\
Sagittarius & $1620 \pm 260$ & $-1060 \pm 600$ & $4050 \pm 540$ & $14.1 \pm 2.7$ & $51.5 \pm 6.4$ & $0.57 \pm 0.09$ & $-85000 \pm 13000$ \\
Cetus & $830 \pm 120$ & $-2000 \pm 390$ & $2350 \pm 330$ & $13.0 \pm 2.1$ & $36.0 \pm 3.8$ & $0.47 \pm 0.04$ & $-99000 \pm 14000$ \\
Wukong/LMS--1 & $300 \pm 110$ & $-650 \pm 110$ & $2540 \pm 180$ & $10.7 \pm 1.0$ & $21.8 \pm 1.9$ & $0.34 \pm 0.06$ & $-120100 \pm 5500$ \\
Orphan & $1240 \pm 380$ & $-3940 \pm 350$ & $2500 \pm 1200$ & $19.8 \pm 5.7$ & $52.9 \pm 5.7$ & $0.48 \pm 0.09$ & $-81000 \pm 11000$ \\
Helmi Stream & $320 \pm 260$ & $-1200 \pm 530$ & $1280 \pm 480$ & $8.4 \pm 0.9$ & $18.6 \pm 4.5$ & $0.37 \pm 0.09$ & $-127000 \pm 21000$ \\
Pal 14 & $7740 \pm 370$ & $410 \pm 770$ & $460 \pm 200$ & $2.2 \pm 1.0$ & $103.1 \pm 1.6$ & $0.96 \pm 0.02$ & $-64260 \pm 590$ \\
Pal 15 & $3460 \pm 240$ & $20 \pm 460$ & $900 \pm 150$ & $1.8 \pm 0.6$ & $48.7 \pm 2.7$ & $0.93 \pm 0.02$ & $-94400 \pm 2500$ \\
\hline
\end{tabular}

\vspace{0.15cm} 
\begin{minipage}{15cm}
\footnotesize
\end{minipage}
\end{table*}

\subsection{Metallicity distribution fitting}
\begin{figure}[t!]
    \centering
    \includegraphics[scale = 0.36]{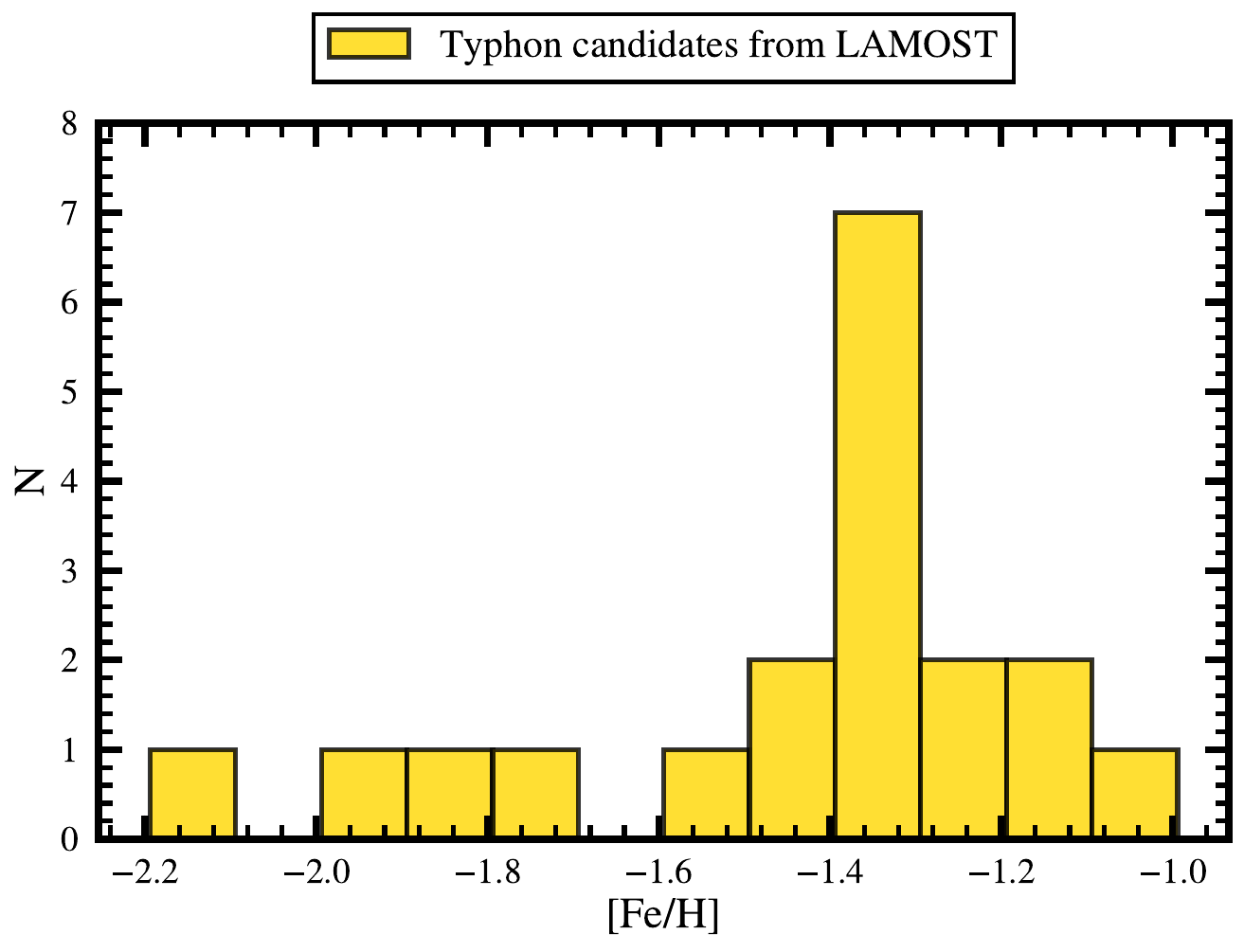}
    \caption{Metallicity distribution of the Typhon stellar stream candidate members identified from the LAMOST survey. The histogram displays the individual stellar metallicities, $\text {[Fe/H]}$, for the 19 candidates selected using kinematic and dynamical criteria, yielding a mean metallicity of $-1.42$\,dex.}
    \label{fig:histogramaLamost}
\end{figure}

To determine the mean metallicity and intrinsic dispersion of the Typhon stellar stream, we model the chemical profile by assuming that the constituent stars follow a Gaussian metallicity distribution function (MDF; see Figure \ref{fig:histogramaLamost}) that is independent of their position along the stream. We constrain these parameters using a Markov Chain Monte Carlo (MCMC) framework. Specifically, we employ the \texttt{EnsembleSampler} from the \texttt{emcee} package \citep{emcee} with 50 walkers. The sampling process consists of a 500-step burn-in phase, followed by a 500-step main production run. Furthermore, to strictly enforce a non-negative metallicity dispersion, we parameterize the model by sampling the logarithm of the dispersion.

Spectroscopic surveys are known to underestimate metallicity uncertainties as they disregard systematic sources of error, such as deviations from the local thermodynamic equilibrium assumption \citep{Soubiran_2022}. To address this, we adopt an empirical approach by testing the inclusion of additional systematic error terms ranging from 0.10\,dex to 0.30\,dex. Applying this test across all four catalogs considered in this work, we find that the intrinsic metallicity dispersion becomes statistically unresolved only when assuming an extremely large additional systematic error approaching 0.3\,dex. Furthermore, to validate the surveys' uncertainties, we examine the abundance spread in globular clusters. This approach relies on the mono-metallic nature of these systems \citep{gratton2019}; because they theoretically possess zero intrinsic dispersion, any observed scatter can be attributed entirely to instrumental or systematic errors. We select NGC~5272 as a reference since it has a similar metallicity ($\mathrm{[Fe/H]} = -1.34$ dex; \citealt{Forbes2010gcs}) to the mean Typhon value. 
By adopting the measured metallicity dispersion of this globular cluster (0.10\,dex) as the additional systematic error term in our calculations, the resulting Gaussian metallicity dispersion for the Typhon stream remains statistically resolved.
\begin{figure}[t!]
    \centering
    \includegraphics[scale = 0.60]{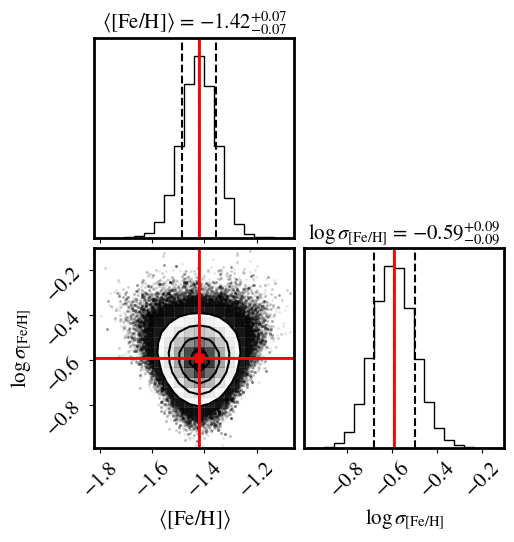}
    \caption{The top panel displays the mean metallicity histogram for the LAMOST sample, which exhibits a Gaussian profile. In the bottom-left panel, the mean metallicity is shown on the horizontal axis against the logarithmic metallicity dispersion on the vertical axis, with solid red lines marking their respective mean values. The bottom-right panel presents the logarithmic metallicity dispersion histogram, which also follows a Gaussian profile. In both histograms, the solid red vertical line represents the mean value, while the black dashed vertical lines indicate the $1\sigma$ uncertainty interval.}
    \label{fig:dispersaoTyphon}
\end{figure}

Figure \ref{fig:dispersaoTyphon} presents the MCMC fit to the MDF from Figure \ref{fig:histogramaLamost} using the LAMOST candidate sample as a representative case. Similar estimates are obtained for all other surveys, and the results are listed in Table \ref{tab:typhon_counts}. Fitting a Gaussian model to the data yields a mean metallicity of $\langle\text{[Fe/H]}\rangle = -1.42\pm0.07$\,dex and a metallicity dispersion of $\sigma_{\text{[Fe/H]}} = 0.26^{+0.06}_{-0.05}$\,dex; note that, for convenience, Figure \ref{fig:dispersaoTyphon} shows the $\log \sigma_{\text{[Fe/H]}}$ posterior. For this calculation, a systematic error term of $0.10$\,dex is added in quadrature to the nominal survey uncertainties for each star to account for potential underestimations related to instrumental limitations, as detailed above. 
Throughout this work, we adopt the results with the LAMOST sample as the main values for Typhon since stellar parameters from this survey have been extensively validated in the literature and have also been used in the original discovery by \citet{Tenachi2022typhon}.

\subsection{Evidence for a disrupted dwarf galaxy}

As depicted in Figure \ref{fig:histogramaLamost}, the Typhon MDF with LAMOST candidates spans a significant [Fe/H] range, from $-2.2$\,dex to $-1.0$\,dex. Although \cite{Tenachi2022typhon} previously utilized this same catalog, our substantially expanded sample allows us to statistically resolve the intrinsic metallicity dispersion of the stream. Even when accounting for both the nominal survey uncertainties and the additional systematic errors derived in our analysis, the observed spread cannot be reconciled with a mono-metallic population. We consistently obtain this same outcome across all catalogs analyzed in this study. 

The DESI sample comprises a good portion of our candidate members ($\sim$65\% of the sample at $\sigma_d/d < 50\%$ and $\sim$35\% at $\sigma_d/d < 20\%$) and yields a lower intrinsic dispersion compared to the other surveys, the chemical spread remains statistically significant. It must be noted that the majority of the identified DESI stars are hot main-sequence turnoff dwarfs.
In this context, \cite{koposov2025} reported that globular clusters with mean metallicities comparable to Typhon typically exhibit dispersions between 0.10\,dex and 0.15\,dex in DESI data. In contrast, the Typhon candidates identified in DESI span a broad metallicity range of $-1.7 \lesssim \text{[Fe/H]} \lesssim -1.0$. This extensive spread is reproduced in SEGUE as well, where members are detected with metallicities approaching $-1.0$\,dex and extending below $-2.7$\,dex. Although the intrinsic metallicity dispersion inferred from DESI is comparatively small, it remains statistically resolved, as it does in every other survey analyzed in this study. The DESI result alone does not exclude a globular cluster progenitor and must be interpreted alongside the larger intrinsic dispersions measured in the other surveys, accounting for the uncertainties and systematic effects affecting each dataset. Taken together, the statistical significance of the metallicity dispersions across all surveys and the larger values measured outside DESI disfavors a mono-metallic globular cluster progenitor and supports a dwarf-galaxy origin for the Typhon stream.



\begin{figure}[t!]
    \centering
    \includegraphics[scale = 0.38]{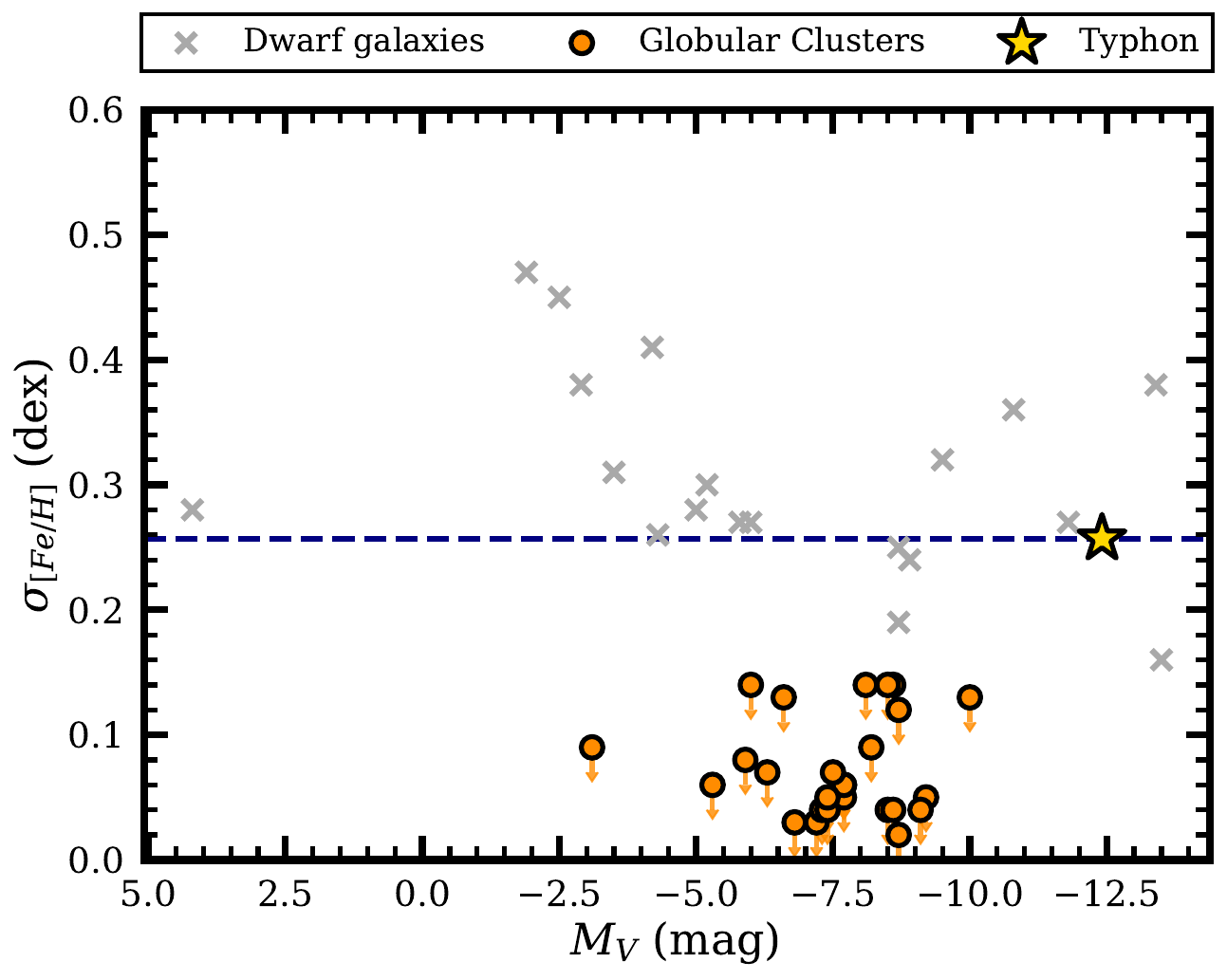}
    \caption{Metallicity dispersion ($\sigma_{[\text{Fe/H}]}$) versus $V$-band absolute magnitude ($M_V$) diagram for Milky Way satellite globular clusters (orange circles) and dwarf spheroidal galaxies (grey crosses), with data compiled from \citet{Geha2026GCs}. Globular clusters occupy the low metallicity dispersion region ($\sigma_{[\text{Fe/H}]} \lesssim 0.1 \text{ dex}$), while dwarf galaxies exhibit larger dispersions. The gold star represents the Typhon stellar stream, and the horizontal blue dashed line highlights its measured metallicity dispersion. Given that this structure was recently accreted by the Milky Way, its absolute magnitude ($M_V \approx -12.41$) is estimated using the relationship between the measured metallicity dispersion and the magnitude of its progenitor dwarf galaxy.}
    \label{fig: dwarfGalaxy}
\end{figure}

It has been made clear that metallicity dispersion serves as a diagnostic tool to distinguish the progenitors of stellar streams in the Galactic halo \citep[e.g.,][]{Li2022S5streams}. Globular clusters are typically mono-metallic and chemically homogeneous populations, exhibiting $\sigma_{[\text{Fe/H}]}$ values close to zero, while dwarf galaxies, due to their extended star formation histories, display significant and measurable metallicity dispersions. This distinction is illustrated in Figure \ref{fig: dwarfGalaxy}, using literature data from \citet{Geha2026GCs}. The value of $\sigma_{\text{[Fe/H]}} = 0.26^{+0.06}_{-0.05}$\,dex obtained for Typhon with LAMOST is substantially different from zero, supporting the dwarf-galaxy nature of Typhon. The derivation of the $V$-band absolute magnitude ($M_V$) is described below. The distribution in Figure \ref{fig: dwarfGalaxy} demonstrates that Typhon, regardless of its progenitor luminosity, can be robustly separated from the globular cluster population. Thus, we conclude that Typhon is likely the first confirmed stellar stream originating from a dwarf galaxy in the outermost Galactic halo. This discovery aligns with expectations from cosmological simulations of Milky Way-mass galaxies, which predict that the outer halo should be occupied by low-mass systems \citep{fatahhi2020simulations, monachesi2019, dsouza2018,Deason2016simulation, Deason2023simulation}.

\subsection{Properties of the Typhon disrupted dwarf galaxy progenitor}
\begin{figure}[t!]
    \centering
    \includegraphics[scale = 0.35]{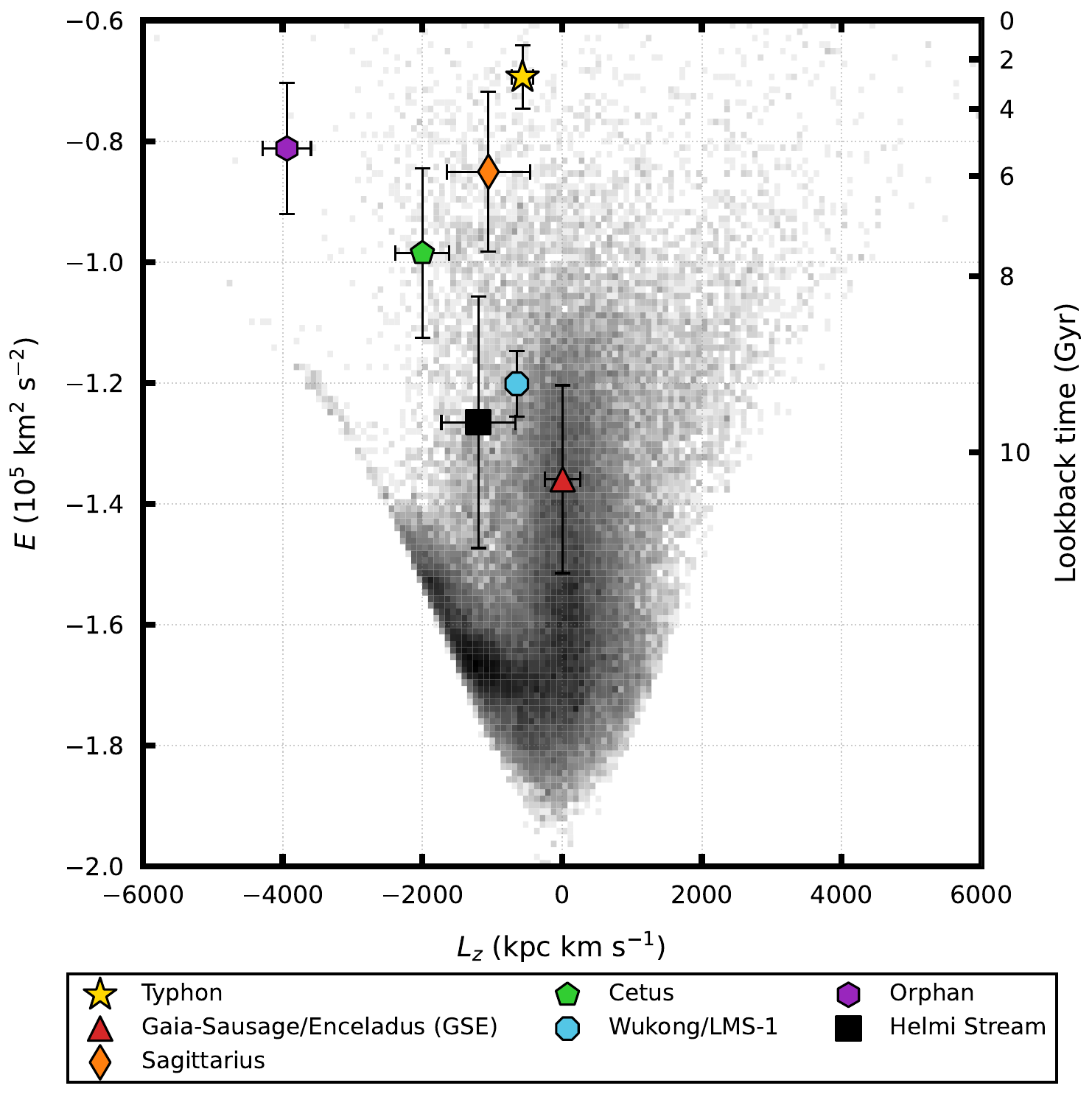}
    \caption{Distribution of orbital energy ($E$) versus $z$-component of angular momentum ($L_z$). The underlying grayscale density map illustrates the phase-space distribution of the background stars. Distinct colored markers represent the mean orbital parameters and corresponding dispersions (error bars) for selected known dwarf galaxy remnants and substructures in the Milky Way halo: Orphan, Wukong/LMS-1, Cetus, Sagittarius, Gaia-Sausage/Enceladus (GSE), and the Helmi Stream. Additionally, the 'X' markers denote the globular clusters Pal 14 and Pal 15. The stellar population associated with the Typhon stream is highlighted by the prominent gold star at the top of the panel. The secondary right vertical axis translates the orbital energy into look-back time (Gyr), providing a proxy for the accretion epochs of these systems.}
    \label{fig:lze}
\end{figure}

To place Typhon's dynamical properties in context, we construct a comparison sample comprising systems of dwarf-galaxy origin selected from \cite{MalhanDwfSelection}.
 Table \ref{tab:parametros_orbitais} summarizes the orbital actions, pericentric and apocentric distances, eccentricities, and specific orbital energies of Typhon and the comparison objects. As previously discussed, axisymmetric potentials admit specific integrals of motion that remain conserved along an orbit. Consequently, stars belonging to the same accreted structure cluster together in the space defined by $L_z$ (or $J_\phi$) and orbital energy ($E$). Figure \ref{fig:lze} displays the $L_z$ vs. $E$ parameter space for various dwarf-galaxy remnants in the Milky Way's halo. 
 

Building on the premise that Typhon originated from a dwarf galaxy, we can use this dynamical space to contextualize its accretion history. Orbital energy serves as a proxy for the accretion epoch of the progenitor system. This relationship is explicitly demonstrated by mapping the phase-space volume occupied by several disrupted substructures detailed by \citet{Naidu2022mzr}. The reference sample spans a broad range of accretion histories, from ancient mergers such as the Gaia-Sausage/Enceladus \citep[GSE, $z \sim 2.0$;][]{belokurov2018, amina2018, Haywood2018,gallart2019, Bonaca2020, Montalban2021, laporte2026} and the Helmi streams \citep[$z \sim 1.0$;][]{helmi1999, koppelman2019} to more recent infall events like Sagittarius \citep[$z \sim 0.6$;][]{Ibata1994, Majewski2004sgr,Laporte2018sgr, Ruiz-Lara2020sag} and the Orphan stream \citep[$z \sim 0.3$;][]{Belokurov2007Orphan, koposov2023orphan}. A distinct correlation emerges where higher orbital energies correspond to more recent accretion events. To calibrate this timeline on the secondary right-side vertical axis of Figure \ref{fig:lze}, we convert the aforementioned accretion redshifts into look-back times assuming a flat $\Lambda$ cold dark matter cosmology \citep{PlanckCollab2020}, yielding values that approximately represent the time elapsed since each progenitor system fell into the Galactic potential.

The kinematics of the Typhon stream, denoted by the gold star symbol, place it in a high-energy regime ($-0.8 \times 10^5 < E/({\rm}\,{\rm km}^2\,{\rm s}^{-2}) < -0.6 \times 10^5$)  with a prograde orbit ($L_z < 0$). Since orbital energy is correlated with accretion time \citep{Rocha2012}, this configuration strongly suggests that Typhon experienced a more recent merger event compared to all the other known disrupted dwarfs in the Milky Way. Consequently, we treat Typhon as the remnant of a recent merger event.

\begin{figure}[t!]
    \centering
    \includegraphics[scale = 0.36]{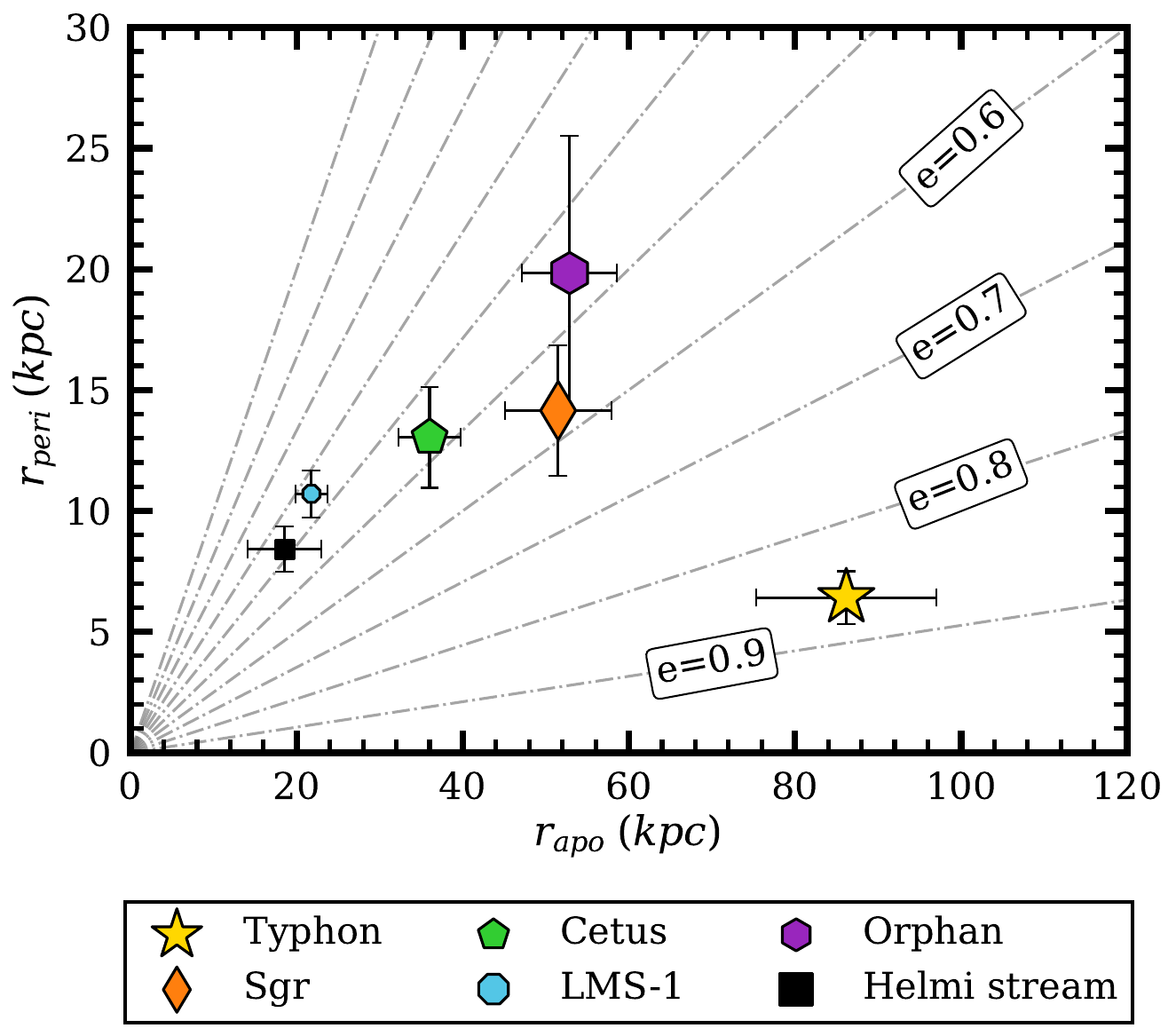}
    \caption{Mean pericenter vs. apocenter distances of known dwarf galaxy remnants in the Milky Way halo (represented by colored symbols). Typhon is shown as an gold star. Dashed lines represent iso-eccentricity curves.}
    \label{fig:ecc}
\end{figure}

Figure \ref{fig:ecc} presents the relationship between the apocentric ($r_{\rm apo}$) and pericentric ($r_{\rm peri}$) radii, highlighting lines of constant orbital eccentricity ($e$), where $e = (r_{\rm apo} - r_{\rm peri})/(r_{\rm apo} + r_{\rm peri})$. The diagram compares the Typhon stream with several known halo dwarf galaxy remnants, including Sagittarius \citep{Johnson2020sgr, Limberg2023sgr}, Cetus \citep{Newberg2009cetus, Yuan2019cetus, Yuan2022cetus}, Wukong/LMS-1 \citep{naidu2020, Yuan2020lms1, Malhan2021lms1, Johnson2022wukong, limberg2024wuk}, Orphan \citep{Koposov2019orphan, koposov2023orphan}, and the Helmi streams \citep{koppelman2019, myeongShards, Limberg2021hstr}. In this parameter space, Typhon exhibits a highly radial orbit. Specifically, it exhibits the largest apocenter distance and the highest eccentricity among all the stellar streams in the comparison sample.

Prior to the discovery of Typhon, no high-eccentricity stellar streams of dwarf galaxy origin had been identified in the outer halo \citep{Malhan2021lms1, Malhan2022atlas}. A possible interpretation for this absence is that progenitors on radial orbits are more rapidly disrupted \citep{Boylan-Kolchin2008, Piatti2019}. This would explain the observation, illustrated in Figure \ref{fig:ecc}, that nearly all surviving stellar streams occupy circular orbits. In contrast, cosmological simulations \citep{shipnora2023,shippnora2025} predict a significantly larger population of high-eccentricity streams. The discovery of Typhon and the strong likelihood of its dwarf-galaxy nature therefore help to alleviate this tension between observations and theory, offering a crucial test case for understanding the survival rates of radial accretion events. These findings also indicate that many more high-eccentricity disrupted dwarfs and streams might be lurking in the outer Galactic halo.

\begin{figure}[t!]
    \centering
    \includegraphics[scale = 0.36]{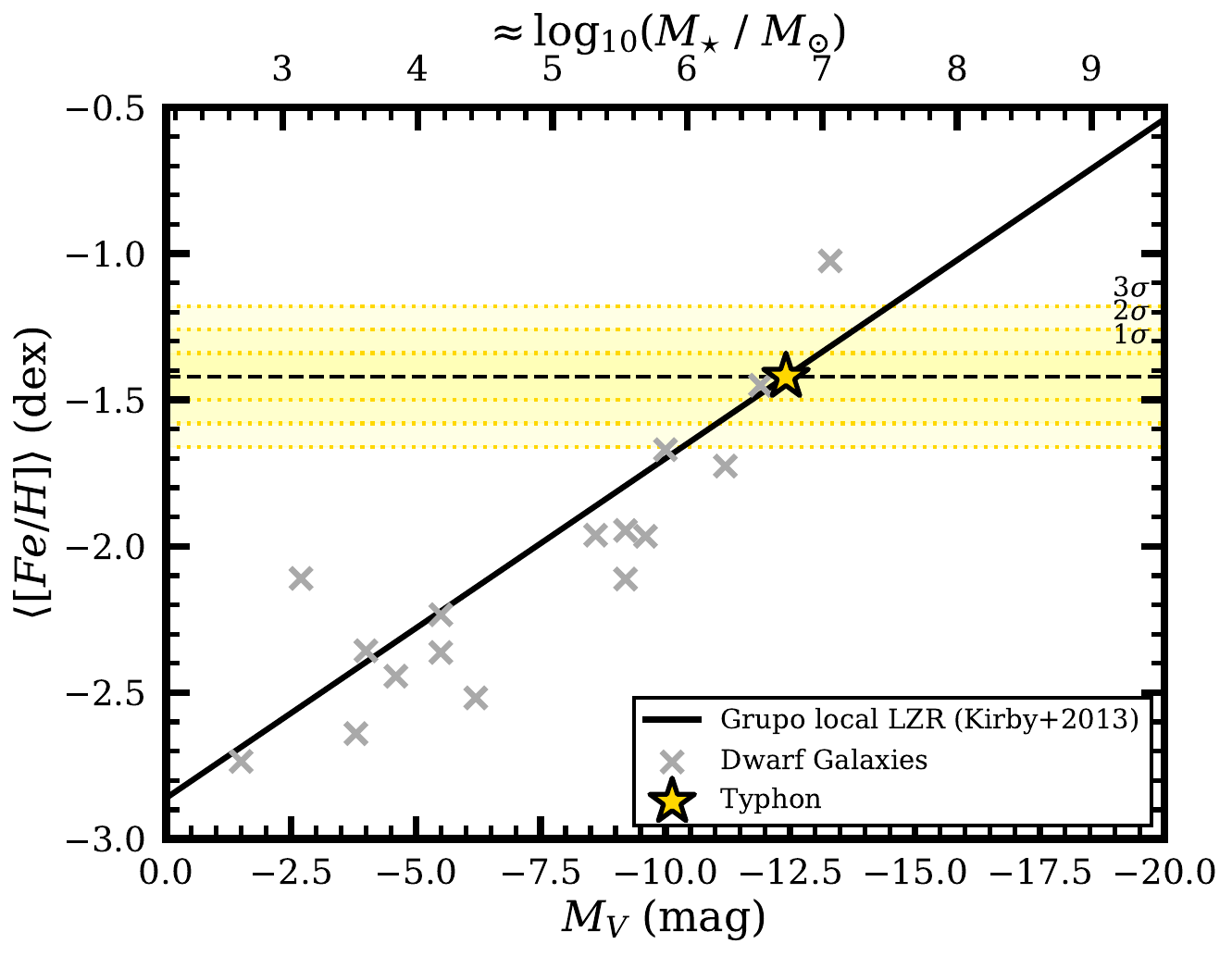}
    \caption{Luminosity$-$Metallicity relation for Local Group dwarf galaxies and the Typhon stellar stream. The grey crosses correspond to Local Group dwarfs from \cite{Kirby2013}, while the gold star denotes Typhon.}
    \label{fig: massaOrigin}
\end{figure}
Figure \ref{fig: massaOrigin} illustrates the luminosity--metallicity relation \citep[LZR;][]{Kirby2013, simon2019}, plotting the mean metallicity ($\langle [\text{Fe/H}] \rangle$) versus the $V$-band absolute magnitude ($M_V$) for Milky Way satellite dwarf galaxies (grey crosses). The magnitude $M_V$ serves as a proxy for luminosity, which is correlated with the approximate stellar mass ($M_\star$) shown on the top axis (assuming a mass-to-light ratio of 2). In this context, the mean metallicity of Typhon stars, derived in this work using LAMOST data, is used to estimate the stellar mass and $V$-band absolute magnitude of its progenitor dwarf galaxy. The solid black line represents the LZR fit for the Local Group, as established by \citet{Kirby2013}. The golden star symbol (Typhon) represents the structure's mean metallicity and estimated $M_V$ of $-$1.42\,dex and $-$12.4\,mag, respectively. Given the recent accretion of the Typhon stellar stream, inferred from its high orbital energy, we estimate the progenitor's original stellar mass using the local (redshift $z=0$) LZR. This yields a stellar mass similar to that of the Sculptor dwarf spheroidal galaxy ($10^6 < M_\star/M_\odot \lesssim 10^7$) \citep{McConnachie2012catalog}.

In \cite{AlexJi2023typhon}, high-resolution spectroscopy of seven stars in the Typhon stellar stream revealed elevated $\alpha$-element abundances, specifically magnesium. Because such an enhancement typically indicates a truncated star formation history preceding the onset of Type Ia supernovae, these authors suggested that Typhon might be the remnant of an ancient accretion event. However, they acknowledged that this interpretation fundamentally conflicts with the dynamical properties of the system, which indicate a recent infall. To reconcile this discrepancy, \citet{AlexJi2023typhon} proposed several scenarios, including the possibility that Typhon represents the high--energy tail of a more massive disrupted galaxy, as such massive progenitors experience greater orbital decay via dynamical friction.

Additionally, an investigation of orbital eccentricity (Figure \ref{fig:ecc}) reveals that while known dwarf galaxies typically exhibit circular orbits, Typhon exhibits a highly radial orbit with $e \sim 0.88$. We speculate that dwarf galaxies on such radial orbits are subject to more rapid tidal disruption due to their deep plunges into the Galactic potential \citep{Boylan-Kolchin2008}. If this orbital characteristic indeed accelerates tidal stripping, the fact that Typhon remains a coherent stellar stream (unlike the phase-mixed GSE) would strongly corroborate a recent accretion scenario. This observation reinforces the possibility that Typhon is the remnant of a disrupted dwarf galaxy captured before efficient disruption mechanisms could fully dissolve the stream.

\citet{Amarante2022gsehalos} performed tailored hydrodynamical simulations of the disruption of the GSE dwarf galaxy around a Milky Way-analog host. By comparing their results with Typhon's orbital properties, we find that GSE can not explain Typhon's origin. This corroborates the results presented in this work, specifically the stellar mass estimates in Figure \ref{fig: massaOrigin}, which indicate that Typhon is too low-mass to be the product of a major merger. Furthermore, since no other known disrupted dwarf galaxy can account for Typhon's combined dynamical, kinematic, and chemical characteristics, we rule out its association with any other known substructure. 

Recently, \citet{globularClusterTyphon} associated two globular clusters (Pal 14 and Pal 15, with phase-space parameters from \citealt{pal14and15parameters}) with the Typhon stream by employing a custom-built orbit integrator with a time-varying potential. We evaluate this claim by integrating the orbits of both clusters using the {\tt AGAMA} package in the \citet{mcmillan2017} axisymmetric potential. By adopting the same solar kinematics as \citet{Tenachi2022typhon} and applying the selection boxes defined in both their study and our present work, we find no kinematic association between these clusters and the Typhon stellar stream. If Pal 14 and Pal 15 were originally hosted by the Typhon dwarf galaxy, they would have fallen into the Milky Way's gravitational potential,  conserving their actions during the tidal disruption process. However, these clusters possess $J_\Phi > 0$, indicating retrograde orbits, unlike the prograde Typhon stream $(J_\Phi < 0)$. The fact that they are counter-rotating relative to the stream corroborates the hypothesis that they are not associated. The resulting absence of associated globular clusters is not inconsistent with our predicted Typhon progenitor stellar mass. \citet{Eadie2022gcs} estimated that only $\sim$50\% of dwarf galaxies within the mass range of the Typhon progenitor ($10^6 < M_\star/M_{\odot} \lesssim 10^7$) host at least one globular cluster
.

\section{Conclusions \& Summary}
\

\label{sec:conc}
The primary objectives of this study are to expand the known membership of the Typhon stellar stream, comprehensively characterize its progenitor, and determine whether it originated from a dwarf galaxy or a globular cluster. To achieve these goals, we develop and apply a selection methodology based on kinematic and dynamical criteria. This approach uses a combination of spectroscopic data from the LAMOST, SEGUE, and DESI surveys, alongside Gaia astrometric and radial-velocity data, metallicities derived from XP spectra, and {\tt StarHorse} distance estimates, with survey-provided distances used for DESI.

The main results of this work can be summarized as follows:

\begin{itemize}
    \item The application of our kinematic selection criteria yields 107 Typhon candidate members with metallicity information for $\sigma_d/d < 50\%$, of which 56 also satisfy $\sigma_d/d < 20\%$. This constitutes the largest such sample in the literature to date, expanding the previously available sample of candidate members with chemical data by more than a factor of ten.

    \item For 19 LAMOST candidates satisfying \(\sigma_d/d<50\%\), the analysis yields a mean metallicity of \(\langle[\mathrm{Fe/H}]\rangle=-1.42\pm0.07\) dex and an intrinsic metallicity dispersion of \(\sigma_{[\mathrm{Fe/H}]}=0.26^{+0.06}_{-0.05}\) dex. This fit includes an additional uncertainty of 0.10 dex added in quadrature to the nominal metallicity uncertainty of each star. 

    \item The identification of three distant candidate members in the DESI survey, located at distances greater than 55\,kpc, with the farthest at 67\,kpc, strongly suggests that Typhon's extremely high apocenter is an intrinsic orbital feature of the system. The previously known sample was restricted to the solar neighborhood ($\lesssim$4\,kpc). Discovering members across these diverse orbital configurations is essential for modeling the kinematics and dynamics of the system. 
    
    \item The resolved metallicity dispersion significantly distinguishes Typhon from mono-metallic populations. This abundance variation strongly supports Typhon's origin as a dwarf galaxy rather than a globular cluster. Furthermore, using the luminosity-metallicity relation of \cite{Kirby2013} with the mean metallicities from the different surveys, we estimate the stellar mass of Typhon's progenitor system to lie in the range $10^6 < M_\star/M_\odot \lesssim 10^7$, comparable to that of the Sculptor dwarf spheroidal galaxy.
    
    \item The kinematics of the Typhon stream place it in a high-energy regime ($-0.8 \times 10^5 < E/({\rm}\,{\rm km}^2\,{\rm s}^{-2}) < -0.6 \times 10^5$) within the \citet{mcmillan2017} axisymmetric potential. Because orbital energy serves as a proxy for the accretion time, this configuration suggests that Typhon experienced a more recent accretion event than other known disrupted Milky Way satellites.

    \item Typhon exhibits a highly radial orbit, featuring the largest apocenter distance and the highest eccentricity ($e \sim 0.88$) among known dwarf-galaxy stellar streams. This extreme orbital eccentricity serves as an additional dynamical indicator of a recent accretion event. Furthermore, because nearly all surviving stellar streams occupy circular orbits, Typhon provides a crucial test case for understanding the survival of radial accretion events in the Galactic halo.

    \item We find no globular clusters that could be dynamically associated with Typhon.
    The previously claimed association of the globular clusters Pal~14 and Pal~15 with Typhon is not supported by our kinematic analysis. 
    
    \end{itemize}

Taken together, our results highlight Typhon as a unique piece of the Milky Way's accretion history puzzle. It offers a rare glimpse of a recently disrupted dwarf galaxy in the outer halo, preserving a dynamical footprint that is quite unlike any other surviving satellite. Future observations from the Rubin Observatory’s Legacy Survey of Space and Time (LSST) survey \citep{ivezic2019lsst} will expand our census of outer halo substructures and likely discover more stellar streams on eccentric orbits.


\software{\texttt{corner} \citep{corner2016}, 
            matplotlib \citep{matplotlib},
            NumPy \citep{numpy},
          SciPy \citep{scipy},
          scikit-learn \citep{scikit-learn},
          TOPCAT \citep{TOPCAT2005}
          }

\acknowledgments

This study was financed, in part, by the São Paulo Research Foundation (FAPESP), Brazil, Process Numbers \#2024/22429-3, \#2026/01108-0,
\#2022/16502-4, \#2024/16510-2,
\#2025/19654-8, \#2024/17850-1, and
\#2020/15245-2. The authors also
acknowledge the partial support from CNPq (Proc. 303816/2022-8, 309431/2026-3, 304535/2026-5, and 192390/2025-2).

This work has made use of data from the European Space Agency (ESA) mission {\it Gaia} (\url{https://www.cosmos.esa.int/gaia}), processed by the {\it Gaia} Data Processing and Analysis Consortium (DPAC, \url{https://www.cosmos.esa.int/web/gaia/dpac/consortium}). Funding for the DPAC has been provided by national institutions, in particular the institutions participating in the {\it Gaia} Multilateral Agreement.

This research has made use of VizieR catalog access tool, CDS, Strasbourg, France. The original description of the VizieR service was published in \citet{VizieR2000}. This work also made use of NASA's Astrophysics Data System Bibliographic Services. 

\bibliography{bibliography}{}
\bibliographystyle{aasjournal}

\appendix

\section{Kinematic Selection Across the Analyzed Surveys}
\label{allSurveyselection}
\begin{figure}[h!]
    \centering
    \includegraphics[scale = 0.36]{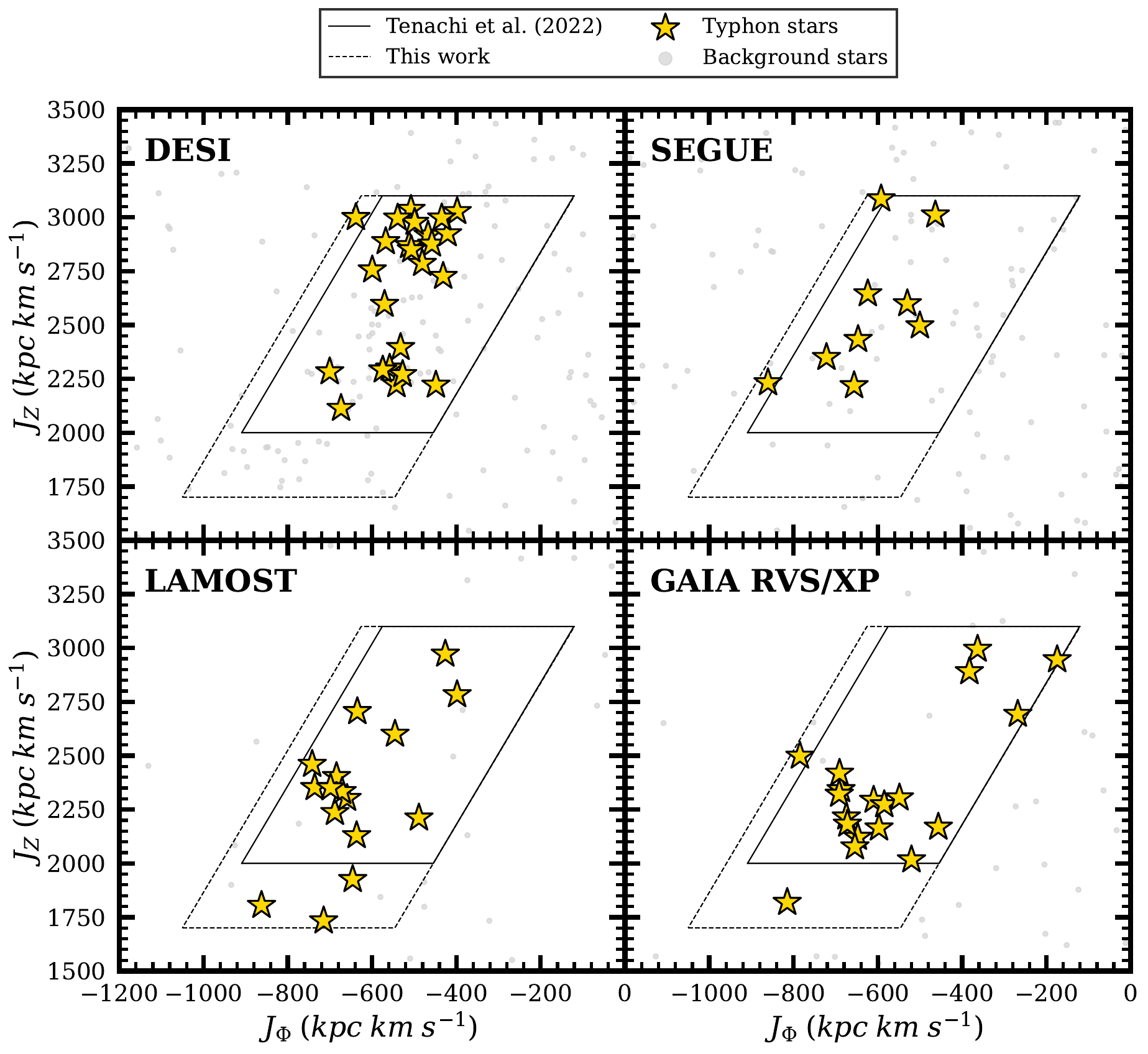}
    \caption{Action-space distribution ($J_z$ versus $J_\Phi$) demonstrating the kinematic selection of Typhon stellar stream members across the DESI, SEGUE, LAMOST, and Gaia RVS/XP surveys. Yellow star symbols represent the final selected Typhon candidates for each respective dataset. Gray points denote background stars with an apocenter $>70\,$kpc. The dashed line illustrates the extended selection region defined in this work, while the solid black line indicates the original selection boundaries established by \citep{Tenachi2022typhon}.}
    \label{fig: appendix1}
\end{figure}

In Figure \ref{fig: appendix1}, we present the action-space distribution ($J_z$ versus $J_\phi$) used for the kinematic selection of Typhon stellar stream members across the DESI, SEGUE, LAMOST, and Gaia RVS/XP surveys (see Section \ref{sec:methods}). The dashed lines illustrate the extended selection region defined in this work, while the solid black lines indicate the original selection boundaries established by \cite{Tenachi2022typhon}. Yellow star symbols represent the final selected Typhon candidates, whereas gray points denote background stars with an apocenter $> 70\text{ kpc}$.

\newpage
\section{Metallicity Distribution Across All Surveys}
\begin{figure}[h!]
    \centering
    \includegraphics[scale = 0.36]{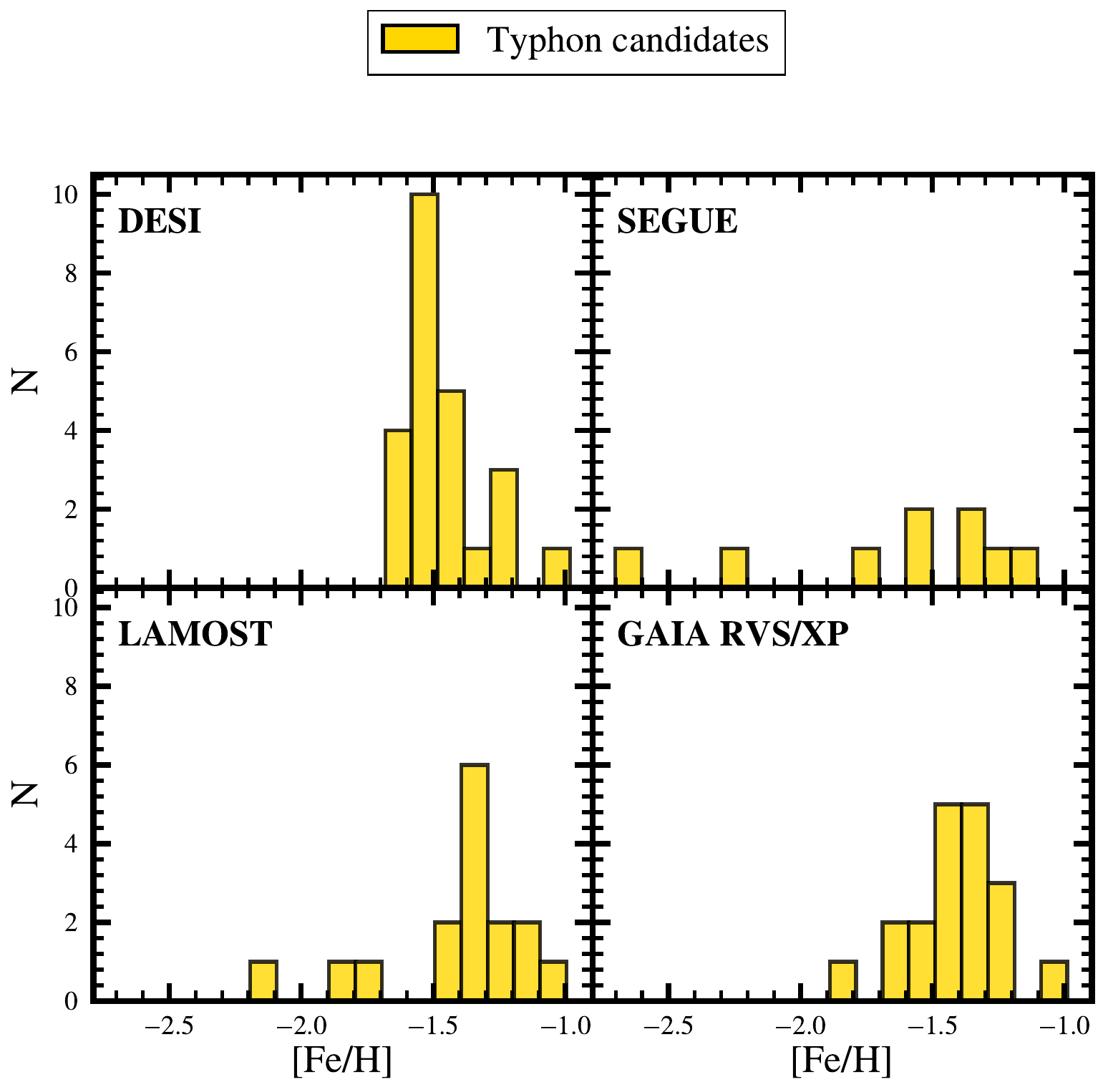}
    \caption{Histograms showing the metallicity ([Fe/H]) distributions for the final Typhon candidates across the four astronomical surveys analyzed in this work. The panels display the distributions for the DESI, SEGUE, LAMOST, and Gaia RVS/XP surveys, respectively.}
    \label{fig: appendix2}
\end{figure}

In Figure \ref{fig: appendix2}, we present the metallicity ([Fe/H]) distributions for the final selected Typhon stellar stream candidates (see Section \ref{sec:res}). The histograms display the chemical abundance spreads across the four astronomical surveys analyzed in this work, with individual panels detailing the resulting distributions for the DESI, SEGUE, LAMOST, and Gaia RVS/XP datasets, respectively.
\label{metalicityAll}

\end{document}